# Quantum-Chaotic Cryptography

R. V. Ramos

rubens.viana@pq.cnpq.br

*Lab. of Quantum Information Technology, Department of Teleinformatic Engineering – Federal University of Ceara - DETI/UFC, C.P. 6007 – Campus do Pici - 60455-970 Fortaleza-Ce, Brazil.*

In this work, it is presented an optical scheme for quantum key distribution employing two synchronized optoelectronic oscillators (OEO) working in the chaotic regime. The produced key depends on the chaotic dynamic and the synchronization between Alice's and Bob's OEOs uses quantum states. An attack on the synchronization signals will disturb the synchronization of the chaotic systems increasing the error rate in the final key.

## 1. Introduction

Nowadays, with the increasing number of network services, data security in optical networks is a crucial issue. Two solutions for data security in optical networks based on physical systems are chaotic cryptography and quantum key distribution (QKD). Chaotic cryptography requires (at least) two synchronized chaotic systems and it takes advantages of pseudo-randomness and high dependence on the parameters values to protect the information [1-6]. There are different optical systems that can be used for chaotic communication. Here it is considered only optoelectronic oscillators producing chaotic light polarization states [7]. Quantum key distribution, by its turn, uses random choices of (non-orthogonal) quantum states and bases of measurement, as well the non-cloning theorem, in order to guarantee the security of the information [8-11]. Some attempts to put chaotic cryptography and QKD to work together have been proposed [12,13], however, in those cases the quantum and the chaotic systems are clearly distinct. In the present work, we propose a setup for quantum key distribution using synchronized optoelectronic oscillators operating in chaotic regime. In this case, the quantum and the chaotic systems are integrated in only one system: the produced key depends on the chaotic dynamic and the synchronization between Alice's and Bob's OEOs uses quantum states. As it will be shown latter, this is a very different kind of quantum key distribution since the quantum states do not carry the key's bits, they are responsible for keeping the synchronization of the chaotic systems.

This work is outlined as follows: In Section 2 the optoelectronic oscillator is reviewed. In Section 3 the optical setup for quantum-chaotic cryptography is presented. Section 5 brings a security analysis. At last, the conclusions are drawn in Section 5.

## 2. Optoelectronic Oscillators Producing Chaotic Light Polarization States

The optoelectronic oscillator here considered is similar to the one that produces chaotic light polarization states described in [7]. Nevertheless, in this work we consider the pulsed regime [13]. The setup is shown in Fig. 1.

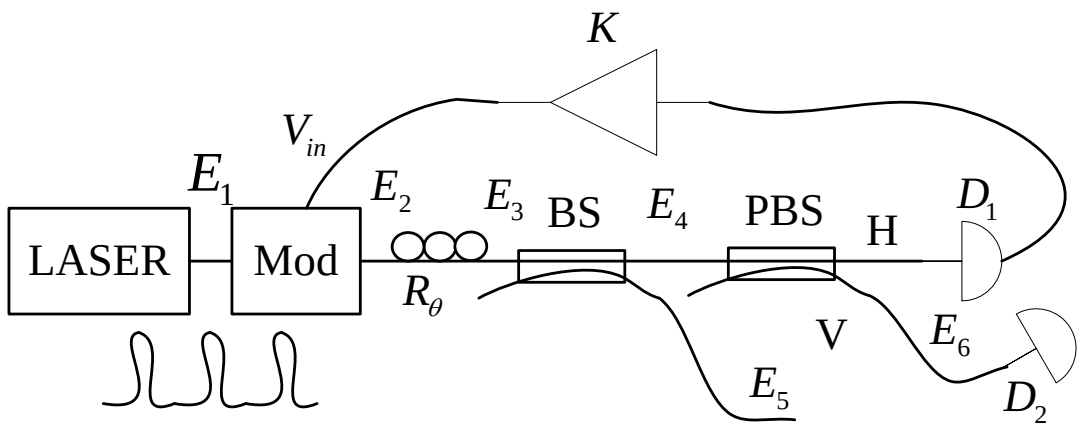

Fig. 1 – Optoelectronic oscillator that produces chaotic light polarization states: BS – beam splitter, $K$ – electrical amplifier, $D_{1,2}$ – photodetectors, PBS – polarizing beam splitter, $R_\theta$ - polarization rotator, MOD – electrooptical polarization modulator and $E_1$-$E_6$ are the electrical fields at the marked positions.

The optoelectronic oscillator is an optical scheme where the light emitted by the laser source is modulated and detected. The photocurrent produced is amplified and used to feed the modulator. Moreover, the time required to the light be detected in $D_1$ and the electrical signal produced to feed the modulator is equal to the time interval between consecutive light pulses generated by the laser. The BS is a balanced beam splitter, PBS is a polarizing beam splitter, $D_1$ and $D_2$ are optical detectors and $K$ is an electrical amplifier. The light polarization states in Fig. 1 and the recurrence equation that describes the dynamic of the OEO are [13]

$$E_1 = \left|\alpha,\alpha\right\rangle_{HV} \tag{1}$$

$$E_2 = \left|\alpha \exp\left(j\left(\frac{\pi}{2}\frac{V_{in}(t)}{V_\pi}+\varphi\right)\right), \alpha \exp\left(-j\left(\frac{\pi}{2}\frac{V_{in}(t)}{V_\pi}+\varphi\right)\right)\right\rangle_{HV} \tag{2}$$

$$E_3 = \left|i\sqrt{2}\alpha \sin\left(\frac{\pi}{2}\frac{V_{in}(t)}{V_\pi}+\varphi\right), \sqrt{2}\alpha \cos\left(\frac{\pi}{2}\frac{V_{in}(t)}{V_\pi}+\varphi\right)\right\rangle_{HV} \tag{3}$$

$$E_4 = \left|i\alpha \sin\left(\frac{\pi}{2}\frac{V_{in}(t)}{V_\pi}+\varphi\right), \alpha \cos\left(\frac{\pi}{2}\frac{V_{in}(t)}{V_\pi}+\varphi\right)\right\rangle_{HV} \tag{4}$$

$$E_5 = \left|-\alpha \sin\left(\frac{\pi}{2}\frac{V_{in}(t)}{V_\pi}+\varphi\right), i\alpha \cos\left(\frac{\pi}{2}\frac{V_{in}(t)}{V_\pi}+\varphi\right)\right\rangle_{HV} \tag{5}$$

$$E_6 = \left|0, i\alpha \cos\left(\frac{\pi}{2}\frac{V_{in}(t)}{V_\pi}+\varphi\right)\right\rangle_{HV} \tag{6}$$

$$V_{in}(t+\tau) = K\left|\alpha\right|^2 \sin^2\left(\frac{\pi}{2}\frac{V_{in}(t)}{V_\pi}+\varphi\right). \tag{7}$$

In (1)-(6) the subscripts $H$ and $V$ mean, respectively, the horizontal and vertical modes. In (7) $\tau$ is the time interval between two consecutives light pulses. The light produce by the laser is linearly polarized in $\pi/4$ ($|\alpha,\alpha\rangle_{HV}$). The optical modulator adds a phase of $\pi V_{in}/(2V_\pi)+\varphi$ in the horizontal component and $-\pi V_{in}/(2V_\pi)-\varphi$ in the vertical component. The voltage $V_{in}$ is the modulating signal while $V_\pi$ is the voltage required to add a $\pi/2$ phase and $\varphi$ is the offset value. The polarization rotator applies a $\pi/4$ rotation in the input state. After the polarization rotator, the optical signal is divided by a balanced beam splitter. One half is the output state, $E_5$, while the second half has its horizontal and vertical components separated by a PBS. The horizontal component is detected in $D_1$ and the photocurrent produced is amplified and used as modulating signal. The vertical component is detected in $D_2$ and its value is used for synchronization purposes. The chaotic behavior is achieved when the values of optical input power ($|\alpha|^2$), $\varphi$ and $V_\pi$ and $K$ (that models the electrical amplifier gain, optical losses and the detector's ($D_1$) efficiency) are properly chosen. The initial value of $V_{in}$ is a random variable that depends on the internal noise of the electronic devices, hence, different OEOs will start with different initial values for $V_{in}$. Due to the recurrence equation (7), $V_{in}$ shows a chaotic behavior that is translated to the polarization state of $E_5$. The Stokes parameter $S_1$ of $E_5$ in Fig. 1 is given by

$$S_1 = -\varepsilon\left|\alpha\right|^2 \cos\left(\pi V_{in}/V_\pi + 2\varphi\right), \tag{8}$$

where $\varepsilon$ is a constant that takes into account details of the polarimeter used to measure $S_1$.

## 3. Quantum-Chaotic Cryptography using Synchronized OEOs

The optical setup for quantum-chaotic cryptography is shown in Fig. 2.

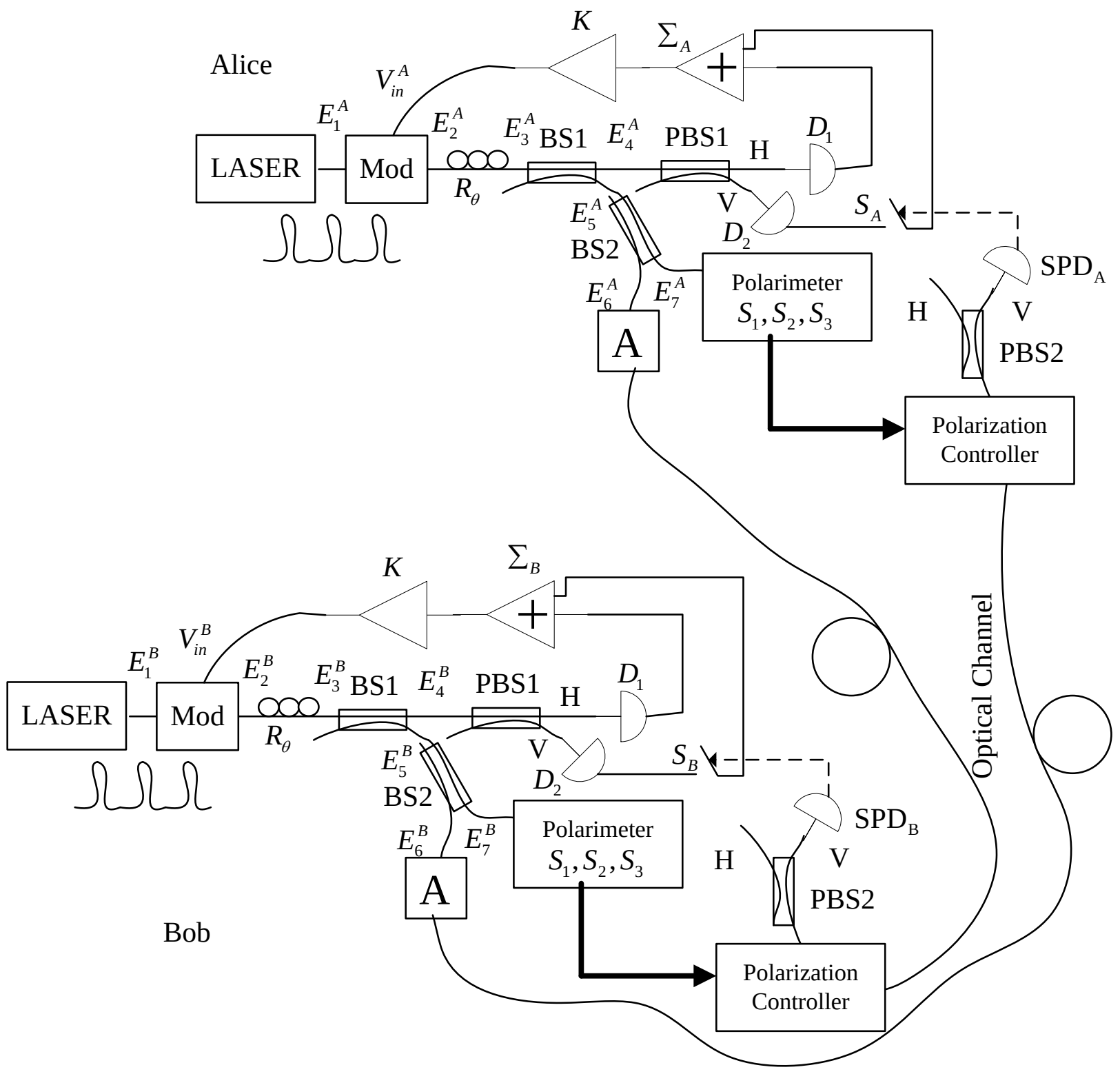

Fig. 2. Optical setup for quantum-chaotic cryptography. $SPD_{A,B}$ – single-photon detector, A – optical attenuator, $S_{A,B}$ – electrical switch and $\Sigma_{A,B}$ - electrical adder.

The system in Fig. 2 works as follows:

1. In Alice (Bob), the output field $E_5^A$ ($E_5^B$) is divided by a beam splitter. One part is strongly attenuated by the optical attenuator *A* and sent to Bob (Alice) through the optical channel. The second part has its polarization measured by a polarimeter. This information is used to control a polarization controller in such way that, if the light coming from Bob (Alice) has the same polarization state, it is NOT guided to the single-photon detector $SPD_{A(B)}$. Hence, Alice (Bob) chooses the basis of measurement (polarization controller and PBS) according to her (his) output light polarization state. Detection in $SPD_{A(B)}$ closes the electrical switch $S_{A(B)}$. In this case the recurrence equation (7) becomes only

$$V_{in}^{A,B}(t+\tau) = K_{A,B}\left|\alpha_{A,B}\right|^2. \quad (9)$$

2. Alice's (Bob's) bit sequence (key) is formed by the discretization of the Stokes parameter $S_1$ of the output optical field $E_7^A(E_7^B)$. In order to obtain a binary sequence from the continuous values of $S_1^{A,B}$, a threshold value $S_{th}$ is defined and, when $S_1^{A,B} < S_{th}$ the bit '0' is obtained otherwise the bit '1' is obtained.

In the scenario of perfect synchronization, both chaotic systems will produce the same light polarizations states at the outputs, $E_7^A = E_7^B$, and the formed key will be the same, since $S_1^A = S_1^B$, implying in a zero bit error rate. On the other hand, a non-perfect synchronization will result in a non-zero error rate: the worse the synchronization the higher is the error rate. In the lack of synchronization, the error rate is around 50%. Hence, in this scheme, the synchronization signals (weak coherent states) are crucial for the security analysis.

As discussed before, the synchronization consists in change the recurrence equation from (7) to (9), what happens when a light pulse coming from Bob (Alice) is detected by Alice (Bob). The probabilities of detection in $SPD_A$ ($q_A$) and $SPD_B$ ($q_B$) are given by

$$q_A \approx p_A t_c |\alpha_B|^2 \left(1-\left[\sin\left(\varphi+\frac{\pi}{2}\frac{V_{in}^A(t)}{V_\pi}\right)\sin\left(\varphi+\frac{\pi}{2}\frac{V_{in}^B(t)}{V_\pi}\right)+\cos\left(\varphi+\frac{\pi}{2}\frac{V_{in}^A(t)}{V_\pi}\right)\cos\left(\varphi+\frac{\pi}{2}\frac{V_{in}^B(t)}{V_\pi}\right)\right]^2\right) \quad (10)$$

$$q_B \approx p_B t_c |\alpha_A|^2 \left(1-\left[\sin\left(\varphi+\frac{\pi}{2}\frac{V_{in}^A(t)}{V_\pi}\right)\sin\left(\varphi+\frac{\pi}{2}\frac{V_{in}^B(t)}{V_\pi}\right)+\cos\left(\varphi+\frac{\pi}{2}\frac{V_{in}^A(t)}{V_\pi}\right)\cos\left(\varphi+\frac{\pi}{2}\frac{V_{in}^B(t)}{V_\pi}\right)\right]^2\right). \quad (11)$$

In (10)-(11), $p_A$ ($p_B$) is the Alices' (Bob') single-photon detector's probability to produce a detectable electrical signal when a photon arrives, $|\alpha_A|^2$ ($|\alpha_B|^2$) is the (lower than 1) mean photon number of the pulse *leaving* Alice (Bob) and $t_c$ is the channel's transmission coefficient. The lower the values of $q_A$ and/or $q_B$ the worse is the synchronization. Thus, like traditional QKD, there is a maximal distance for which the error rate is still acceptable.

A simulation with 40,000 points and with the following parameters values was performed: $\varepsilon$ = 0.01, $|\alpha|^2$ = 100 (for Alice and Bob), $\varphi = \pi/4$, $V_\pi$ = 1V, $K_A = K_B$ = 0.0133, $p_A t_c|\alpha_B|^2 = p_B t_c|\alpha_A|^2$ = 0.03, $V_{in}^A(t=0) = 0.1$ and $V_{in}^B(t=0) = 0.2$ (the two chaotic systems are equal but they start with different initial conditions for $V_{in}$). Furthermore, the OEOs are working without synchronism during the first 100 time slots. One can see part of it in Fig. 3 the almost perfect synchronism between the Stokes parameters $S_1^A$ (+) and $S_1^B$ (o) ($n$ is the slot time number). The error rate was 5.3% in a bit string with 40,000 bits.

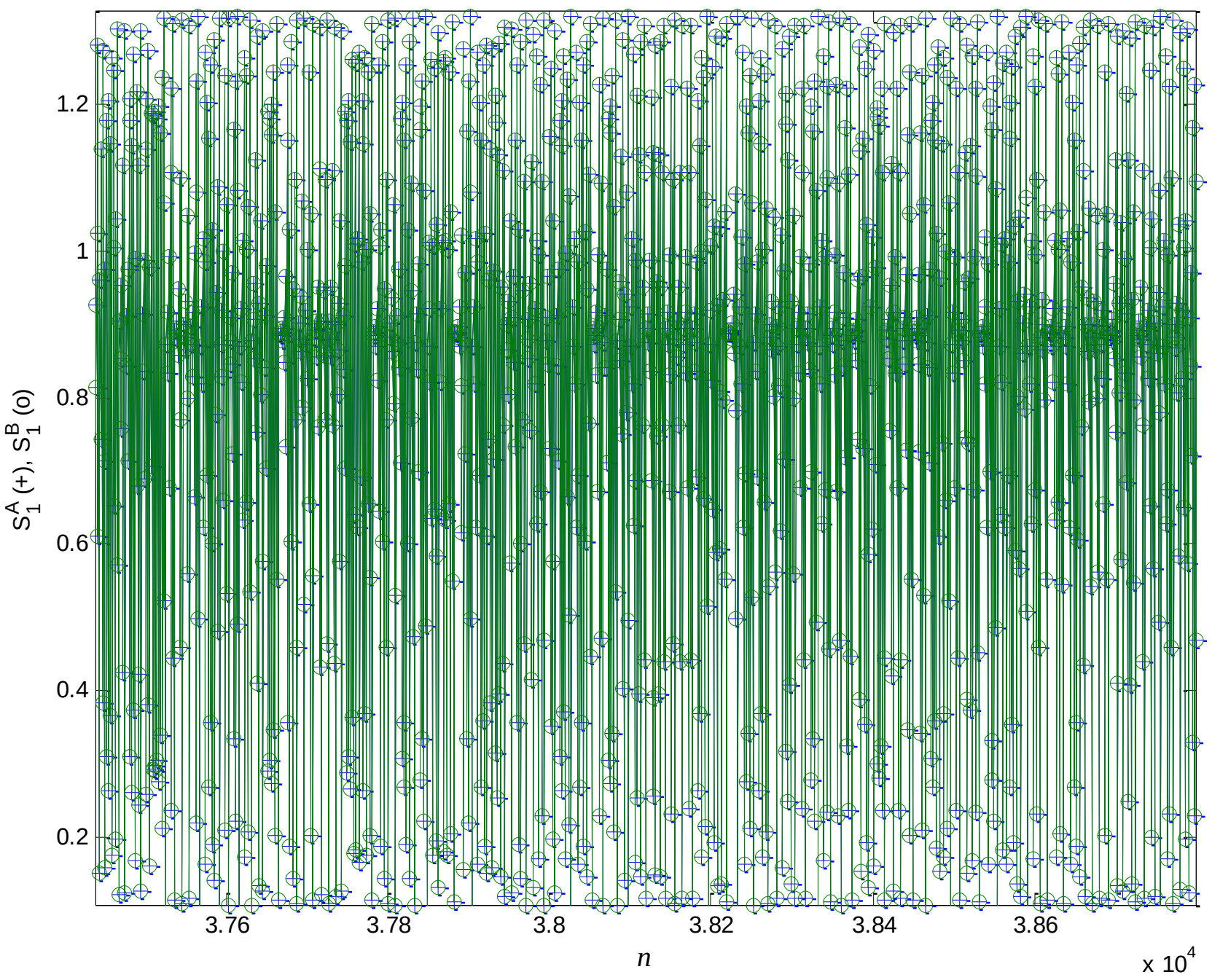

Fig. 3 - $S_1^A$ and $S_1^B$ versus $n$ (almost perfect synchronism).

Another simulation using $K_A$ = 0.0133, $K_B$ = 0.0132, $p_A t_c|\alpha_B|^2 = p_B t_c|\alpha_A|^2$ = 0.83 (the other parameters have the same values used before) can be seen in Fig. 4. In this figure one can see the effect of the

mismatch parameter noise [14] ($K_A \neq K_B$) on the synchronism. The error rate was 23.8% in a bit string with 40,000 bits.

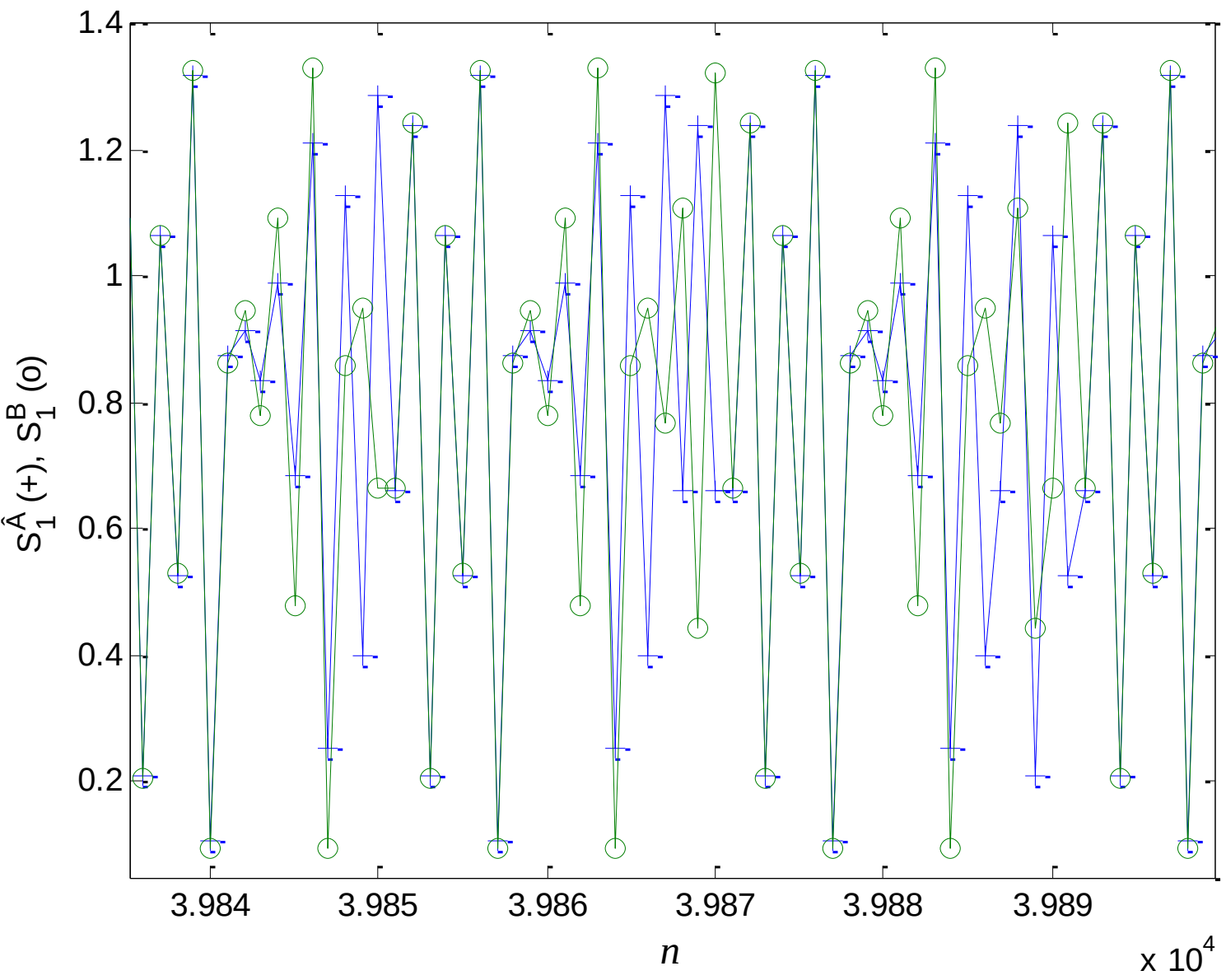

Fig. 4 - $S_1^A$ and $S_1^B$ versus $n$ (non-perfect synchronism).

## 4. Security Analysis

In order to get the bits of the key, a spy must have an OEO synchronized with Alice's and Bob's OEOs. Hence, the eavesdropper will have to attack the quantum signals used in the synchronization and to guess correctly the OEOs parameters values used by Alice and Bob, otherwise the mismatch noise will not permit the synchronization. Thus, as expected, the security of the quantum-chaotic cryptography is guaranteed by quantum and chaotic rules.

During an attack, the eavesdropper cannot decrease the mean photon number of the light pulses that arrives at Alice and Bob (for example, increasing the loss of the channel by inserting a beam splitter) once this would make worse the synchronization, increasing the bit error rate. On the other hand, the eavesdropper could make an attack that would increase detection in Alice's and Bob's single-photon detectors. For example, strong light pulses could be sent to Alice and Bob. However, in this case the dynamic of the non-linear system could change from chaotic to quasi-periodic, for example, indicating the attack. This can be seen in Fig. 5, where the discrete Fourier transforms of $S_1^A$ for $p_A t_c|\alpha_B|^2$ =0.73 and $p_A t_c|\alpha_B|^2$ = 1000 (situation where $SPD_A$ always has detection) are shown. One can clearly see the presence of narrow peaks, showing a quasi-periodic behavior, when this kind of attack happened. In order to avoid this kind of Trojan horse in which strong pulses are sent, an optical detector can be plugged to the non-used PBS2 output (at Alice and Bob) in order to check if strong light pulses are being sent by an eavesdropper.

At last, the eavesdropper can just measure the synchronization signals and to send weak coherent states to Alice and Bob in polarization states that are in agreement with the measurement results. Without knowing in which basis to measure (the light polarization of the pulses sent by Alice and Bob follows a chaotic dynamic, hence, it is a continuous variables), the eavesdropper will introduce a noise that here is modeled by a uniform random variable. Numerical simulations showed an error around 50% in this case.

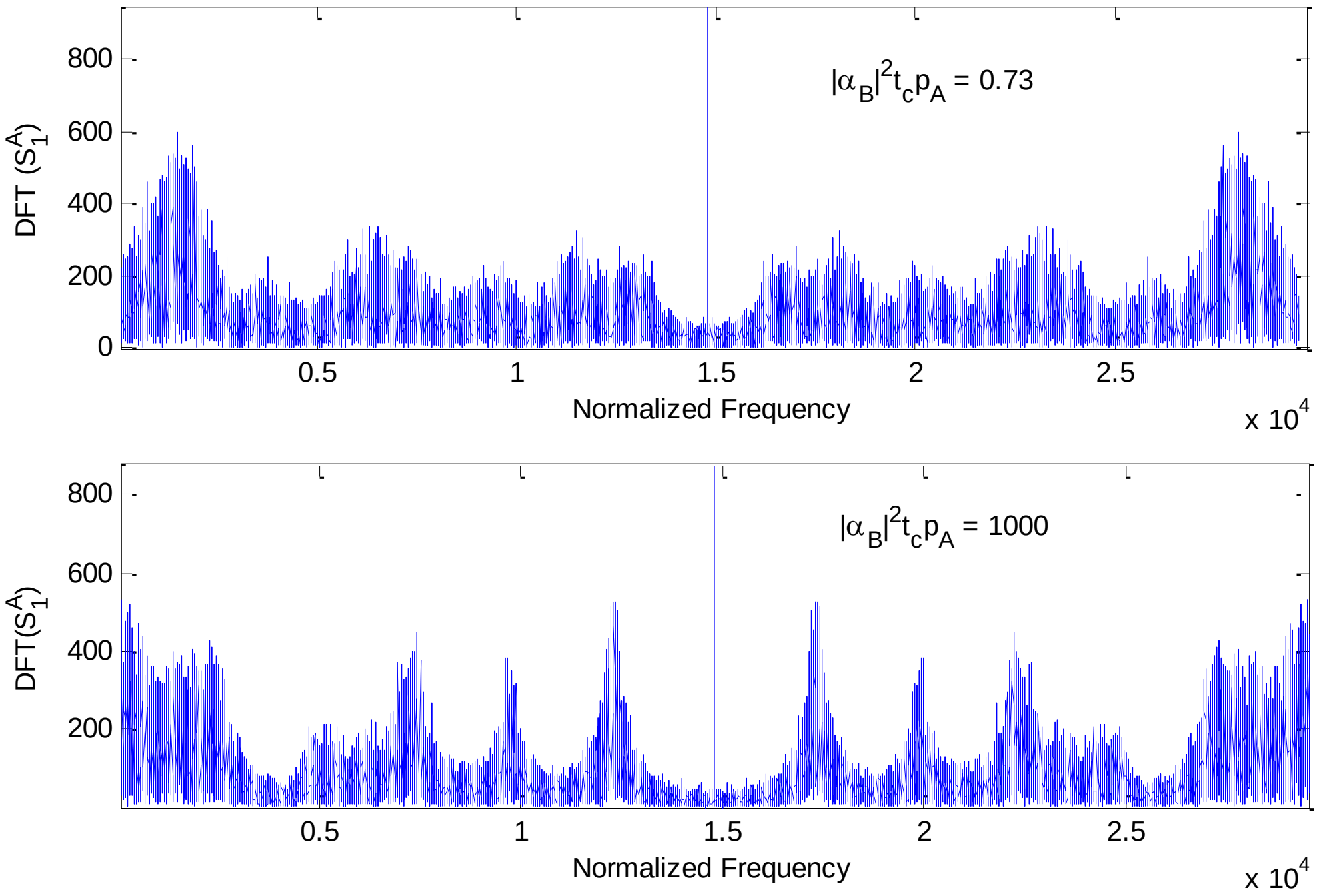

Fig. 5 - Discrete Fourier transform of of $S_1^A$ for $p_A t_c|\alpha_B|^2$ =0.73 and $p_A t_c|\alpha_B|^2$ = 1000.

## 5. Conclusions

The present work showed an optical setup for quantum-chaotic cryptography. Its main characteristics are:

1) The set of quantum states (light polarization state) used is continuous since $V_{in}$ is a continuous variable.

2) The key information is not carried by the quantum states. These are responsible for synchronization of Alice's and Bob's OEOs. They bits of the key depend on the chaotic dynamic of the Stokes parameter $S_1$. They are obtained by measuring high mean photon number coherent states.

3) There is no bases reconciliation stage.

4) Since the synchronization signal is not always present (because of the low mean photon number used for them), the error rate is very sensible to the mismatch noise, that is, when the parameters values are not exactly the same. In the lack of synchronism, the mismatch noise makes impossible two OEOs operating in the chaotic regime to follow the same orbits in the strange attractor. In other words, the larger the mismatch parameter noise, the larger must be the values of $p_{A,B}t_c|\alpha_{A,B}|^2$ in order to still obtain an acceptable error rate.

5) The security is guaranteed by quantum and chaotic rules. The eavesdropper can attack the synchronization signal however, since weak coherent states are used, the polarization is a continuous variable and the OEOs' parameters values are not known, it cannot use this information to try to reconstruct the chaotic behavior of Alice's and Bob's OEOs.

## Acknowledgements

This work was supported by the Brazilian agency CNPq via Grant no. 307062/2014-7. Also, this work was performed as part of the Brazilian National Institute of Science and Technology for Quantum Information.